\documentclass[iop]{emulateapj}

\def\lapp{\ifmmode\stackrel{<}{_{\sim}}\else$\stackrel{<}{_{\sim}}$\fi}
\def\gapp{\ifmmode\stackrel{>}{_{\sim}}\else$\stackrel{>}{_{\sim}}$\fi}
\usepackage{multirow}
\usepackage{color}
\usepackage{amsmath}
\usepackage{soul}
\usepackage{hyperref}

\begin{document}

\title{Signatures of Intra-binary shock emission in the black widow pulsar binary PSR~J2241$-$5236}

\author{
Hongjun An\altaffilmark{1,*}, Roger W. Romani\altaffilmark{2}, and
Matthew Kerr\altaffilmark{3}
\\
{\small $^1$Department of Astronomy and Space Science, Chungbuk National University, Cheongju, 28644, Republic of Korea}\\
{\small $^2$Department of Physics/KIPAC, Stanford University, Stanford, CA 94305-4060, USA}\\
{\small $^3$Space Science Division, Naval Research Laboratory, Washington, DC 20375-5352, USA}\\
}
\altaffiliation{$^*$hjan@chungbuk.ac.kr}

\begin{abstract}
	We report on high-energy properties of the black widow pulsar
PSR~J2241$-$5236 in the X-ray and the {\it Fermi}-LAT (GeV gamma-ray) bands.
In the LAT band, the phase-averaged gamma-ray light curve shows orbital modulation
below $\sim$1\,GeV with a chance probability ($p$)
monotonically decreasing with time to $p\sim 10^{-5}$.
The peak of the light curve is near the superior
conjunction of the pulsar (binary phase $\phi_{\rm B}\approx 0.25$).
We attribute the modulation to the intra-binary shock (IBS) emission
and search for IBS signatures in the archival X-ray data. We find
that the X-ray spectral fit requires a non-thermal component,
which implies a possible IBS origin of
the X-rays. We discuss our observations in the context of IBS scenarios.
\end{abstract}


\section{Introduction}

Many millisecond pulsars (MSPs) have been discovered by {\it Fermi} \citep[LAT;][]{fermimission},
especially so-called spider binaries where the pulsar spin-down power is heating and evaporating the companion.
At present, in the Galactic disk these discoveries include $\sim$15 `redback' (RB) systems with 
stellar-mass companions, and $\sim$20 `black widow/tidarren' (BW) systems with sub-stellar 
companion masses; similar spider binaries are found in the globular clusters.\footnote{https://apatruno.wordpress.com/about/millisecond-pulsar-catalogue/}
The gamma-ray emission from these binaries is dominated by 
$L_{\gamma}\approx10^{33}-10^{35}\rm \ erg\ s^{-1}$ \citep[][]{fermi2PC} pulsed
emission from the MSP magnetosphere, which is independent of orbital phase.

In the X-ray band the sources are less luminous with $L_X\approx 10^{30}-10^{33}\rm \ erg\ s^{-1}$,
and a hard $\Gamma_X\le 1.5$ spectrum \citep[][]{rmgr+15};
the broad 0.3--79\,keV X-ray emission is largely unpulsed and often modulated with binary phase \citep[e.g. ][]{apg15},
especially in the RB.
This unpulsed emission is believed to be produced by an intra-binary shock (IBS) between 
winds from the pulsar and the heated companion. The shocked pulsar wind develops a mildly relativistic
flow in a shell surrounding the source with lower wind momentum flux, so that the shock
synchrotron and inverse Compton radiation is beamed.  When viewed at moderate to high 
orbital inclination this beaming produces orbital modulation \citep[][]{rs16,ar17,whvb+17}.
One sees a double peak \citep[e.g., PSR~J2129$-$0429, PSR~J2215+5135;][]{rmgr+15,grmc+14},
or a broad hump as the line of sight cuts through or grazes the cone of IBS emission.
In the gamma-ray band only a few systems seem to show orbital modulation
and these are the lower-mass BW sources
\citep[e.g., PSR~J1959+2048, PSR~J1311$-$3430;][]{wtch+12,xw15,arjk+17}.
In these sources, gamma-ray IBS emission is believed to be produced by synchrotron
and/or inverse-Compton processes \citep[e.g.,][]{b14}.
The emission competes with the strongly dominant pulsed gamma rays
and so detections are challenging. Improved measurements and
additional examples are needed to probe the origin of this orbital modulation.

        Here, we report on the high-energy variability of the BW system
PSR~J2241$-$5236. The nearby ($DM$-estimated distance of $\sim$0.5\,kpc) 2.2-ms pulsar is
in a 3.5-h orbit with a $M$$>$0.012M$_\odot$ companion \citep[][]{kjrf+11}. With
its short spin and orbital periods and the proximity, this source is of special 
interest. We have thus used {\it Fermi} Large Area Telescope (LAT)
and archival X-ray observations in order to search for high-energy
orbital variability and probe its IBS.

\section{Observational Data and Analysis}
\label{sec:sec2}
\subsection{Fermi-LAT data}
\label{sec:sec2_1}

	PSR~J2241$-$5236 (J2241 hereafter) is bright in gamma rays,
and the pulsations are easily detected \citep[][]{kjrf+11}.
Although no orbital modulation was reported in that work,
with 9-yr LAT exposure and Pass 8 reprocessing \citep[][]{fermiP8}
we can now make a sensitive measurement of the orbital light curve.
We downloaded the Pass 8 data from the
{\it Fermi} Science Support Center collected between
2008 August 04 and 2017 Aug. 10 with an $R=20^\circ$ aperture in the
60\,MeV--500\,GeV energy band.
We further selected Source class events with Front/Back event type using
zenith angles $\le90^\circ$,
and analyzed the data with the {\it Fermi}-LAT Science Tools {\tt v10r0p5}
along with P8R2\_V6 instrument response functions.\footnote{https://fermi.gsfc.nasa.gov/ssc}

	Optimal extraction of the source photons uses energy-dependent probability
weights. To obtain these we first measure the source spectrum, using a 
binned likelihood analysis in the 100\,MeV--300\,GeV band (35 energy bins), including
corrections for energy dispersion\footnote{https://fermi.gsfc.nasa.gov/ssc/data/analysis/documentation/\\Pass8\_edisp\_usage.html}
using the Python packages provided with the {\it Fermi} Science tools.
We used the 3FGL models \citep[][]{fermi3fgl} in which J2241 emission
is modeled with a power-law-exponential-cutoff model (PLEXP, $\sim$$13\sigma$ preferred over a simple power-law model),
$dN/dE=N_0(E/E_0)^{-\Gamma_{\rm 1}}e^{-(E/E_{\rm c})^b}$, where $b$
is held fixed at 1; if $b$ is free, the result is consistent with 1 and the fit
does not improve.

	In the fit, we vary parameters for bright sources ($\gapp$5$\sigma$) in the aperture
and the normalizations of the diffuse emission models \citep[{\tt gll\_iem\_v06.fits} and
{\tt iso\_P8R2\_SOURCE\_V6\_v06.txt};][]{fermigllv06,fermiiso}. 
Because the best-fit parameters of J2241 can depend on the background from
these sources, we ran tests varying the size of the region of interest
(RoI;  $R=5^\circ$ and $R=15^\circ$) and allowing various numbers of sources 
within the RoI to adjust their parameters from the assumed 3FGL values.
In all cases the J2241 parameters were consistent with those of
the original fit and a scatter an order of magnitude smaller than the statistical errors.
The best-fit parameters for J2241
are $\Gamma_{\rm 1}=1.33\pm0.05\pm0.03$, $E_{\rm c}=2.9\pm0.2\pm0.1$\,GeV,
with 100\,MeV--300\,GeV flux of $3.15\pm0.12\pm0.10\times10^{-8}\rm \ phs\ cm^{-2}\ s^{-1}$.
Here the first error range is statistical and the second is the dominant systematic error
associated with uncertainty in the Galactic diffuse background model and in the LAT effective area.
The 100\,MeV--300\,GeV energy flux is
$3.31\pm0.08\times 10^{-11}\rm \ erg\ cm^{-2}\ s^{-1}$, and the isotropic gamma-ray luminosity
for an assumed distance of 0.5\,kpc is $9.9\pm0.2\times 10^{32}\rm \ erg\ s^{-1}$.
These parameters are consistent with previous LAT spectral analyses \citep{fermi2PC,fermi3fgl}.

	Because IBS emission is commonly modeled with a power law different from the
pulsar model, we check to see if there is evidence for IBS emission in the gamma-ray spectrum.
We fit the LAT spectrum with the addition of a power-law (IBS) component.
A $\Gamma=2.3\pm0.1$ power-law component is detected with the test statistic ($TS$) of $193$
with changes in the above PLEXP parameters,
and the fit improves when including the power-law model ($p\sim0.007$).
However, this could be due to the gamma-ray source in the 3FHL catalog
at angular distance $\sim$0.2$^\circ$ \citep[3FHL~J2240.3$-$5240$, \Gamma_{\rm 3FHL}=2.3\pm0.4$;][]{3fhl};
adding this source to our fits makes an additional IBS PL component
statistically unnecessary.
Below we use PLEXP as our baseline model (no PL).
Including IBS PL emission does not change our conclusions on
orbital modulation significantly.

\begin{figure*}
\centering
\begin{tabular}{cc}
\includegraphics[width=3.3 in]{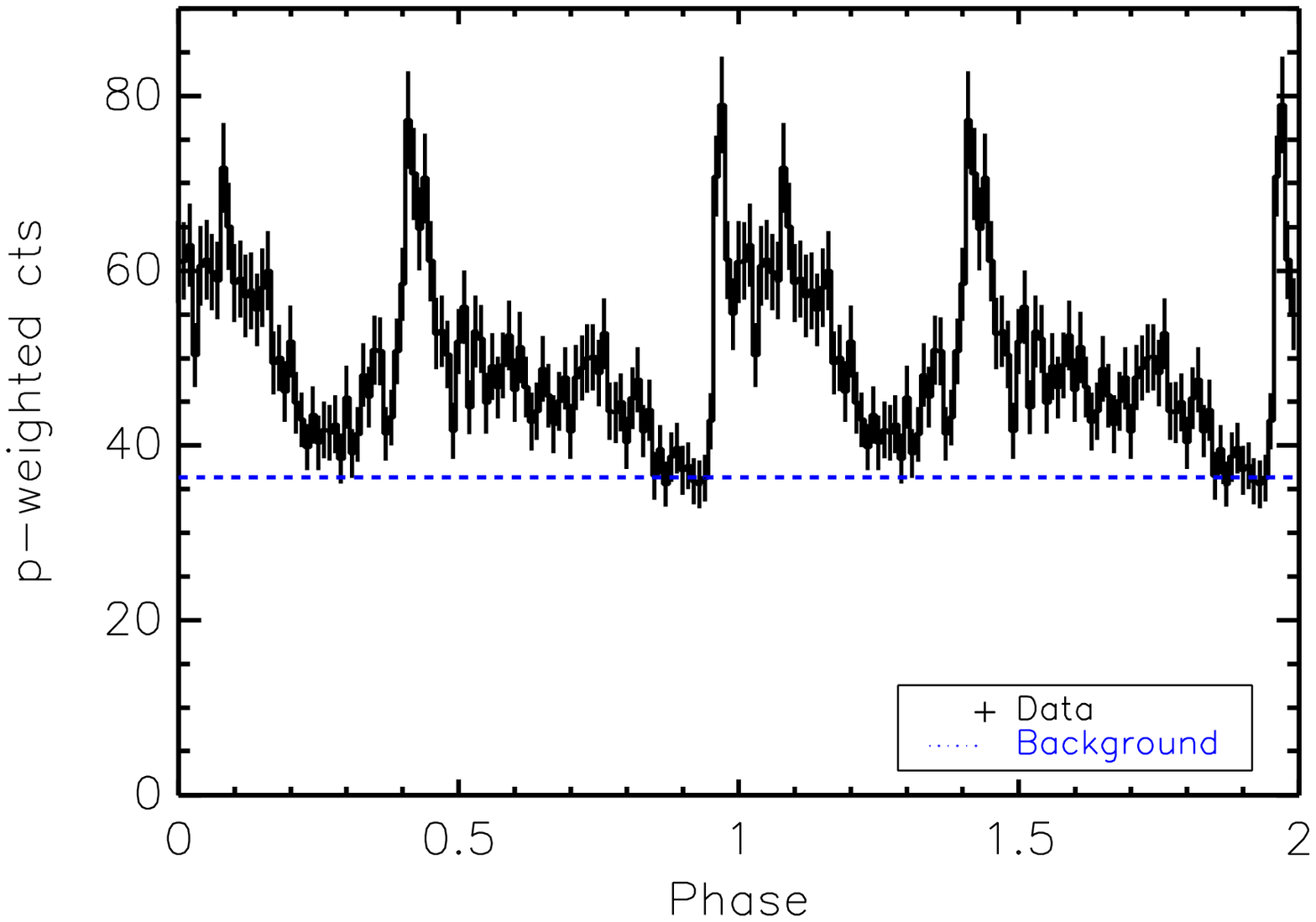} &
\includegraphics[width=3.3 in]{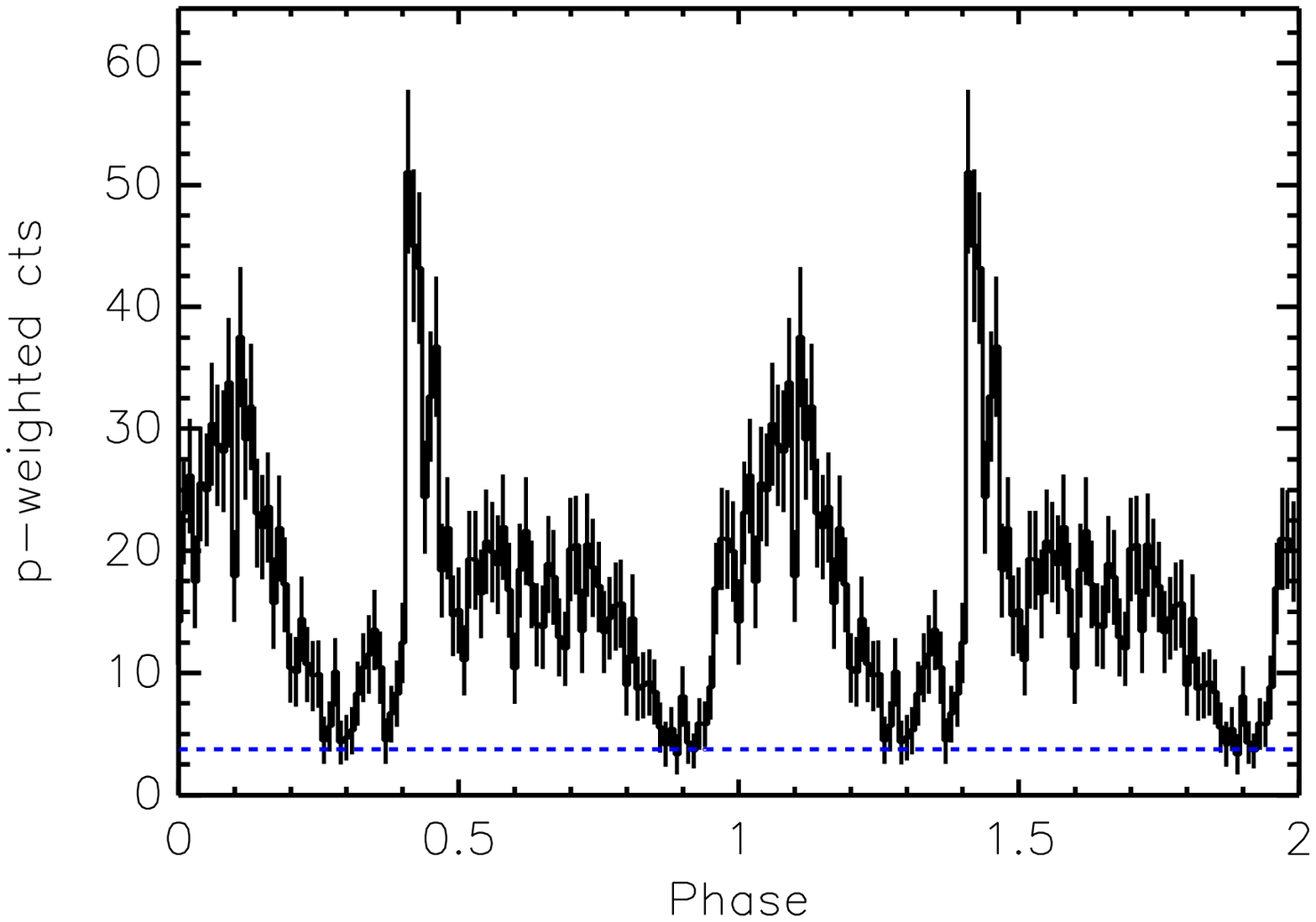} \\
\end{tabular}
\figcaption{Gamma-ray pulse profiles (100 phase bins) in the 60\,MeV--1\,GeV (left)
and 1\,GeV--500\,GeV (right) bands for $R<5^\circ$ extraction.
The background level, as determined using the method of \citet{fermi2PC}, is shown in blue.
\label{fig:fig1}
}
\vspace{0mm}
\end{figure*}

	We have re-examined the pulsar timing solution over the 9-yr LAT data set. 
We start from the existing timing solution\footnote{http://www.slac.stanford.edu/~kerrm/fermi\_pulsar\_timing/J22\\41-5236/html/J2241-5236\_54683\_56587\_chol.par}
and folded the events using {\tt tempo2}. Here, we weight each event with the probability of the
event originating from J2241. The probability weight $w_i$ \citep[][]{k11} is calculated using {\tt gtsrcprob}
of the LAT Science Tools which uses the energy spectrum measured above and the position of the event.
Source pulsations are clearly detected and the profile is similar to the
previous one \citep[][]{kjrf+11}. We, however, notice that the pulse phase drifts slightly
at late times ($\Delta\phi\approx 0.02$), and so we derive an improved timing solution.

	We created a spin ephemeris spanning the length of our data set 
by maximizing the unbinned likelihood
$\log\mathcal{L}=\sum_i \log [w_i f(\phi(\lambda, t_i)) + 1-w_i]$, with
$f$ an analytic model of the pulse shape and $\phi(\lambda,t_i)$ the phase assigned at time $t_i$ by the timing
model with parameters $\lambda$. We evaluated $\phi$ using
the PINT software package\footnote{{https://github.com/nanograv/PINT}} \citep[][]{lrdr+18}, and
found that optimizing $\nu$, $\dot{\nu}$, the position and proper motion, $P_B$,
$A1$, and $T_0$ were sufficient to produce a sharp pulse profile over the full interval.
The best-fit parameters and their 1-$\sigma$ confidence intervals
from the posterior distribution are reported in Table~\ref{ta:ta1},
and the resulting pulse profiles appear in Figure~\ref{fig:fig1}.

\newcommand{\markap}{\tablenotemark{a}}
\begin{table}
\centering
\caption{Timing Parameters for PSR~J2241$-$5236.}
\label{ta:ta1}
\begin{tabular}{ c c }
\hline
\hline
\rule{0pt}{3ex}
RA ($\alpha$, J2000)              & $22^h41^m42.016452^s$(15)\\ 
\rule{0pt}{3ex}
DEC ($\delta$, J2000)            & $-52^\circ  36' 36.2098''$(25)\\
\rule{0pt}{3ex}
Epoch (MJD)             & 55044.15587 \\
\hline
\rule{0pt}{3ex}
$\nu$ (s$^{-1}$) & 457.31015684738(2) \\
\rule{0pt}{3ex}
$\dot \nu$ (s$^{-2}$) & $-$1.4423(2)$\times 10^{-15}$ \\
\rule{0pt}{3ex}
TZRMJD & 56547.72358340209 \\ 
\rule{0pt}{3ex}
Binary model  & BT\markap \\
$P_{\rm B}$  (day) & 0.1456722372(3) \\
$A1$  ($lt$-$s$) & 0.025791(3) \\
$e$ & 0 \\
$T_{0}$& 55044.157905(4) \\
PMRA  ($\dot{\alpha}\cos\delta$, mas\,$yr^{-1}$) & 19.4(4) \\
PMDEC ($\dot{\delta}$, mas\,$yr^{-1}$) & $-$6.1(5) \\
\hline
\hspace{0.5 mm}
\end{tabular}\\
\footnotesize{{\bf Notes.} 1$\sigma$ uncertainties are shown in brackets, and parameters without
the uncertainty are held fixed.\\}
$^{\rm a}${See \citet{ehm06} for the timing model definition.}\\
\end{table}

\begin{figure*}
\centering
\hspace{-5.0 mm}
\begin{tabular}{cc}
\includegraphics[width=3.22 in]{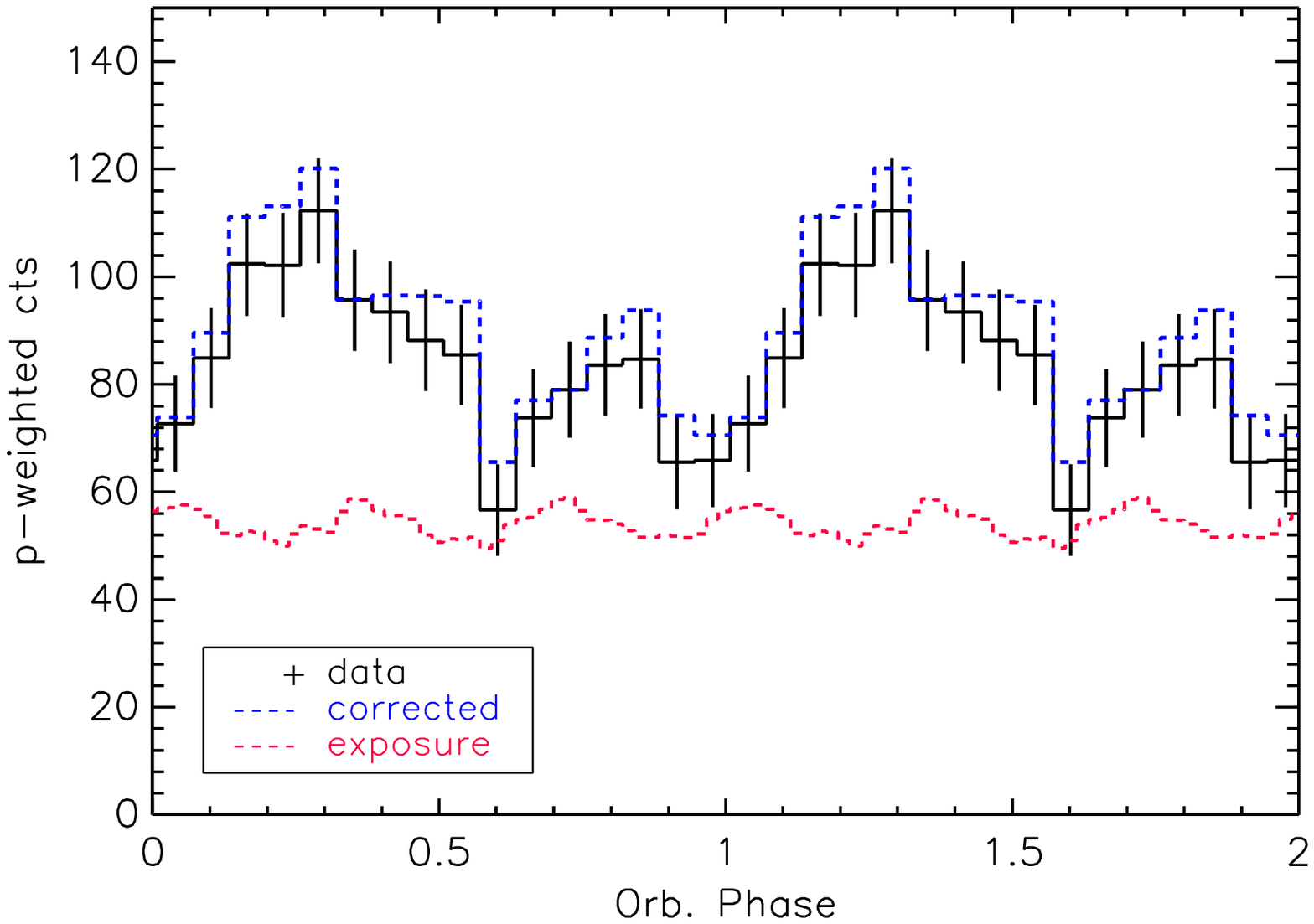} &
\includegraphics[width=3.22 in]{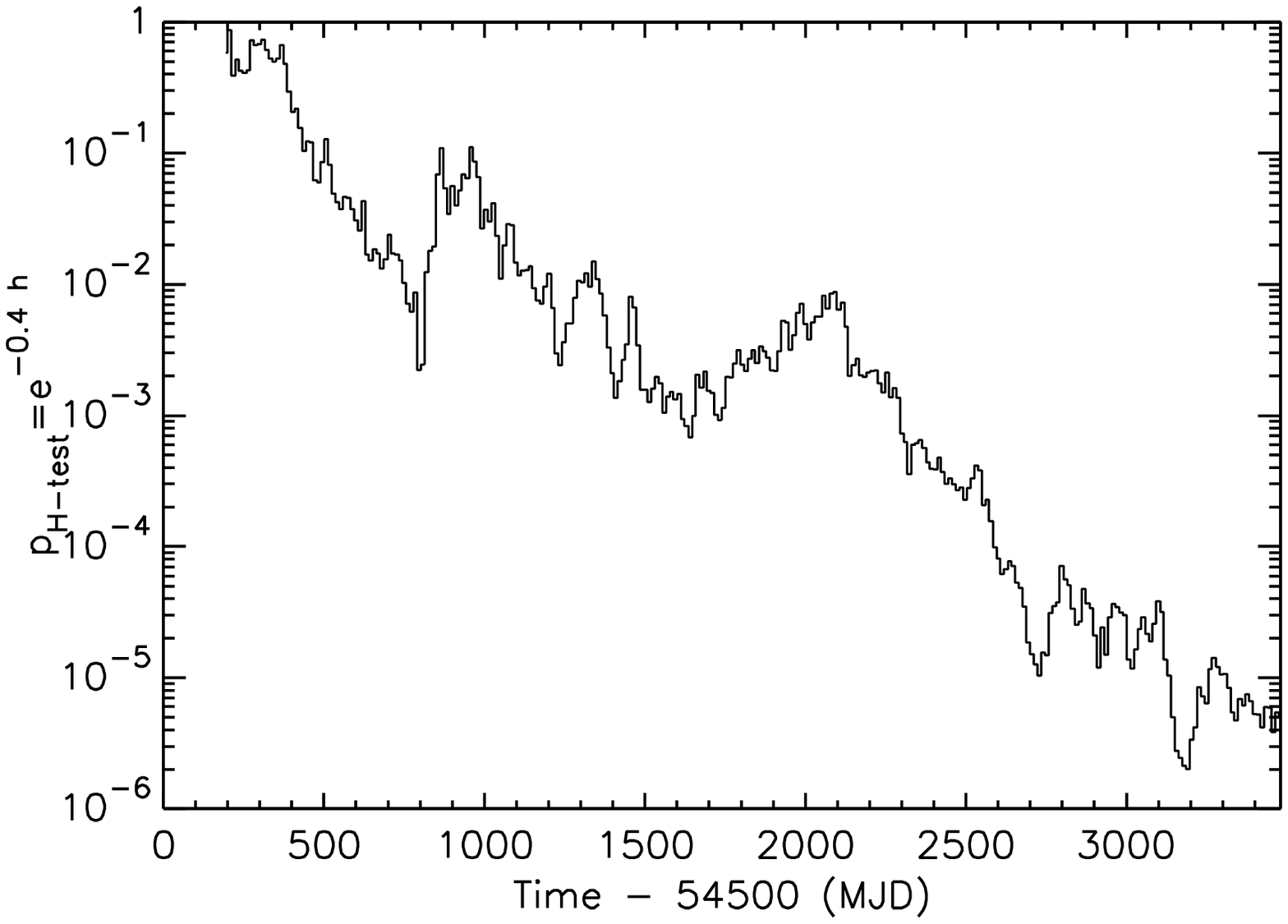} \\
\end{tabular}
\figcaption{{\it Left}: Phase-averaged probability-weighted orbital light curve 
(background subtracted) in the low-energy 0.06--1\,GeV band (black histogram, $H\sim30$). 
Time-resolved exposure (arbitrary scale) is calculated using the 30-s binned spacecraft 
file and is folded with the timing solution in Table~\ref{ta:ta1} (red dotted curve).
The exposure-corrected light curve (blue histogram) has the same relative uncertainties 
as the black histogram.
{\it Right}: Time-cumulative $H$-test \citep[e.g.,][]{arjk+17} results for J2241. Events
are extracted using a $R=5^\circ$ aperture in the 0.06--1\,GeV band, and the probabilities
that the folded light curve is consistent with a constant level are plotted.
\label{fig:fig2}
}
\vspace{0mm}
\end{figure*}

	Next, we investigate orbital variability in the gamma-ray band. Motivated 
by our study of the pulsar PSR~J1311$-$3430, we constructed light curves by
probability-weighted aperture photometry ($R=5^\circ$)
in low- and high-energy bands as well as the full 0.06--500\,GeV band using all spin phases.
We find that the low-energy (0.06--1\,GeV) light curve shows significant orbital modulation 
with a probability of constant signal $\sim10^{-5}$ using the
$H$ test \citep[$H\approx30$ for three harmonics summed, Figure~\ref{fig:fig2} left;][]{drs89, k11}
while the high-energy light curve does not.
The results are not very sensitive to energy cuts, although for the lowest energy
photons ($E=60-700$\,MeV) the constant probability is as low as $\sim10^{-7}$.
As expected for a real signal, the cumulative-time $H$ test \citep[e.g.,][]{arjk+17} 
for the $<1\,$GeV band shows a monotonic decrease of the null hypothesis probability, 
reaching $\sim10^{-5}$ (Fig.~\ref{fig:fig2} right). The full band data also show growing significance, but
the probability reaches only $p\sim 4\times 10^{-3}$. The high-energy band shows no significant 
modulation ($p\sim 0.1$).

Note that the peak of the orbital light curve is near pulsar superior conjunction ($\phi_B=0.25$)
and another smaller one is near $\phi_{\rm B}=0.75$.
In order to see if the smaller peak is statistically required, we fit the
light curve with a single or a double Gaussian function and perform an $f$-test.
We find that the double Gaussian fit is not better than the single Gaussian fit with
$f$-test probability $\gapp 0.1$.

	We checked the phase-dependent exposure variation in our data set.
This is shown by the red dashed line in Figure~\ref{fig:fig2} left. This does
not contribute to the low-energy orbital signal as the amplitude is
small ($\sim$3\%) and does not align with the observed modulation. Correcting for
this small exposure variation gives the blue light curve. Simulated exposures
following this red curve were also used to compute variability induced 
by the $H$ statistic. For this `systematic' error estimate we again find
a similar significance ($p\sim10^{-5}$) for our measured $H$.

	The bright, highly variable, blazar 3FGL~J2329.3$-$4955 (a.k.a, PKS~2326$-$502)
lies $d$$\sim$8$^\circ$ from J2241. Although probability weighting decreases its
contribution to the J2241 signal, one might worry that its substantial
LAT flaring activity 
\citep[e.g.,][]{d10}\footnote{https://fermi.gsfc.nasa.gov/ssc/data/access/lat/msl\_lc/source\\/PS\_2326-502}
might induce modulation to the J2241 light curve. We checked this by folding
the photons from the nearby sources 3FGL~J2235.3$-$4835 and 3FGL J2207.8$-$5345
on J2241's orbital period. No significant modulation is seen, so the blazar 
contribution is insignificant, and we conclude that the J2241 orbital modulation is intrinsic.

	As the orbital modulation seems to be real, we check to see if the spectrum
also varies orbitally.
We first fit the orbital maximum ($\phi_{\rm B}=$0--0.6) and minimum ($\phi_{\rm B}=$0.6--1)
spectra with the 3FGL model. In the fits, we varied the pulsar parameters but held
the other parameters fixed at the best values obtained above. We find that
the maximum spectrum is softer
($\Gamma_{\rm psr}^{\rm max}=1.42\pm0.06\pm0.09$
{\it vs.} $\Gamma_{\rm psr}^{\rm min}=1.22\pm 0.08\pm0.10$; $\Delta \Gamma=0.2\pm0.1$)
with larger flux.
Note that our systematic uncertainties shift both indices similarly;
the $\Gamma$ difference is largely unaffected.
Assuming that the minimum spectrum best represents the orbitally constant pulsar spectrum,
we hold the pulsar parameters fixed at the best values for the minimum spectrum,
add a power-law (IBS) component to the models, and fit the maximum and minimum spectra.
In this case, the minimum spectrum does not require a power-law component ($TS\approx0$)
while the maximum spectrum does ($TS\approx 60$). The model with a $\Gamma_{\rm PL}\approx2.3$
power-law component ($F_{0.1-300\rm GeV}=7.6\pm1.2\times10^{-9}\rm \ phs\ cm^{-2}\ s^{-1}$, orbital average)
fits the maximum spectrum better ($p=0.007$). The results are similar
if we let the amplitude of the pulsar component vary in order to
mitigate the possibility of pulsar emission varying. This suggests that the orbitally-varying component
has a soft power-law spectrum as expected in IBS models.

	Although the steady increase in significance suggests that the modulation
is not episodic, J2241 itself has
a variability index of 61 in the 3FGL catalog \citep[][]{fermi3fgl}, implying 
$\sim$10\% significance for variability on a month time scale. To test this possibility, we performed
a likelihood analysis to extract source fluxes in the 0.1--300\,GeV band in 1\,Ms time bins
for J2241 and three variable and bright sources in the aperture
(3FGL~J2207.8$-$5345, 3FGL~J2235.3$-$4835, 3FGL~J2329.3$-$4955)
while holding the other parameters fixed at the phase-averaged values.
We then constructed a long-term light curve with the time bins having $TS>9$.
The probability that the J2241 light curve is consistent with a constant is 0.06,
not very different from the 3FGL value, implying no significant long-term
variability.

\subsection{X-ray data and Analysis}
\label{sec:sec2_2}

	Since a number of spider binaries (BWs and RBs), especially the RB systems, show strong
orbital modulation in the X-ray band, we examined the 20\,ks of archival ACIS
data collected on 2009 August 30 with the {\it Chandra} observatory. In an earlier
analysis of this data set, \citet{kjrf+11} modeled the source spectrum with blackbodies,
as might be expected from the hot polar cap of the pulsar. However,
some BWs and RBs show a power-law spectral component in the X-ray band, indicating emission from the IBS, so
a re-examination of the ACIS spectrum seems appropriate.

	We extracted counts in a $R=2''$ aperture centered at the source position
and computed the spectral response files using the {\tt specextract} tool of CIAO~4.8.
The background spectrum was extracted using an annular aperture with $R_{\rm in}=5''$
and $R_{\rm out}=10''$. We then grouped the spectrum to have 1 event per spectral bin
and fit the spectrum in the 0.3--10\,keV band in {\tt XSPEC}~12.9.0n
using the $l$ statistic \citep[][]{l92}. The results are shown in Table~\ref{ta:ta2}.

\newcommand{\marka}{\tablenotemark{a}}
\newcommand{\markb}{\tablenotemark{b}}
\newcommand{\markc}{\tablenotemark{c}}
\newcommand{\markd}{\tablenotemark{d}}
\begin{table}[t]
\vspace{-0.0in}
\begin{center}
\caption{X-ray data fit results}
\label{ta:ta2}
\vspace{-0.05in}
\scriptsize{
\begin{tabular}{ccccccc} \hline\hline
model\marka  &   $kT_1$   &  $kT_2/\Gamma$  &  $F_{\rm X}$\markb & $l$/dof\\
             &   (keV)    &  (keV/$\cdots$) &         &           \\ \hline
BB           &  $0.27\pm0.02$ & $\cdots$ & $2.6\pm0.3$  & 59/64 \\ 
2BB          &  $0.10\pm0.06$ & $0.32\pm0.05$ & $2.9^{+0.9}_{-0.5}$ & 56/62 \\ 
PL           &  $\cdots$  & $2.5\pm0.2$ & $3.9\pm0.5$   & 48/64  \\ 
BB+PL        &  $0.24\pm0.03$ & $1.5$\markc & $3.8^{+1.0}_{-0.7}$\markd & 45/63 \\ \hline \hline
\end{tabular}}
\end{center}
\vspace{-0.5 mm}
\footnotesize{
{\bf Notes.} $N_{\rm H}$ is held fixed at $1.2\times 10^{20}\ {\rm cm^{-2}}$.\\}
$^{\rm a}${BB: blackbody, PL: power law.}\\
$^{\rm b}${0.5--10\,keV flux in units of $10^{-14}\rm\ erg\ cm^{-2}\ s^{-1}$.}\\
$^{\rm c}${Frozen.}\\
$^{\rm d}$ {Total flux in the 0.5--10\,keV band in units of $10^{-14}\rm\ erg\ cm^{-2}\ s^{-1}$.
The BB flux is $2.0\pm0.4$ in the same units.}\\
\end{table}

	In all fits we held the hydrogen column density $N_{\rm H}$ fixed at $1.21\times10^{-20}\rm \ cm^{-2}$,
the value used by \citet{kjrf+11}. Our single and double blackbody fits give
temperatures and fluxes consistent with their earlier results. However, we find that
a simple power-law model provides a better fit \citep[$\Delta l = 11$;][]{l92}
corresponding to $p\approx0.004$ \citep[$\sim$3$\sigma$;][]{aic74}, although the spectral index
is relatively soft at $\Gamma=2.5\pm 0.2$. Since other spider binaries with non-thermal
X-ray emission have a relatively hard spectrum \citep[e.g., PSR~J1959+2048;][]{hkth+12},
we also considered the possibility that the spectrum is composite, with a softer
thermal component. Unfortunately, with only 82 events in the source region we cannot
fit for multiple spectral parameters, so we fixed the hard power law at a typical
$\Gamma=1.5$, and find a plausible $kT_1=0.24$\,keV for the black body spectral component,
accounting for about half the flux. As expected $l$ is even less for this composite
model (Table~\ref{ta:ta2}), but the decrease is insufficient to require an extra component, so we must
await deeper X-ray observations to see if the non-thermal emission is required.

	We also folded the ACIS counts on our derived orbital ephemeris. After 
exposure correction, the resulting light curve is flat, with a K-S probability
of constancy of 87\%. Again deeper X-ray integrations will be needed
to probe the spectrum and modulation of this component.

\section{Discussion and Conclusions}
\label{sec:sec3}

	We have analyzed high-energy data for the BW pulsar J2241 collected
with {\it Chandra} and {\it Fermi}.
We find that in the low-energy gamma-ray band the source exhibits orbital
modulation peaking near the superior conjunction of the pulsar.
Motivated by this temporal signature of IBS in the LAT data,
we investigate the gamma-ray spectrum for the IBS power-law emission, and
find that there seems to be an underlying power-law emission but the
emission could be from a known 3FHL source; no clear signature of IBS
is found in the orbital-phase-averaged LAT spectrum of J2241.
However, we find that the orbital hump spectrum is softer
than the dip spectrum and that a $\Gamma\approx 2.3$ power-law component
is detected in the hump when subtracting the dip spectrum.
This suggests that a part of the power-law emission may be from the IBS.
In the X-ray band we see no strong orbital modulation, although the
sensitivity is low due to limited source counts. However, we do find
that the X-ray spectral fits are improved if a power-law component
is included, suggesting the presence of non-thermal IBS emission that
might be probed with a deeper observation.

	In pulsar binaries, it is believed that X-rays
and low-energy gamma rays are produced by synchrotron emission of shock-accelerated
electrons in the IBS. When the pulsar-wind momentum flux dominates
that of the companion, the IBS wraps around the secondary \citep[e.g.,][]{rs16,whvb+17}.
Shock-accelerated electrons experience acceleration due to
adiabatic expansion as they flow along this surface \citep[][]{bkk08}, leading 
to modest bulk Lorentz factors. Thus the IBS emission is beamed 
along the tangent of the shock. For a pulsar-wind dominated shock we then expect
IBS emission centered around phase $\phi_B\approx 0.25$, at pulsar
superior conjunction, as for J2241 (Figure~\ref{fig:fig2} left).
With the Earth line of sight tangent to the IBS we would get a single
peak; higher inclinations can produce a double peak. Thus
our sub-GeV modulation is consistent with a Doppler-boosted emission
from an IBS.

	There is also a suggestion of a local maximum in Figure~\ref{fig:fig2} (left) at $\phi=0.75$.
The significance is very low ($f$-test $p\gapp0.1$), so it is likely just a statistical fluctuation.
However, in principle, peaks at both phases can be accommodated if the companion wind strength
fluctuates so that the momentum ratio $\beta$ of the pulsar and the companion winds
varies about 1. Perhaps for J2241 the
typical state has $\beta<1$ so that the IBS wraps around the companion star (giving
a peak at phase 0.25) while occasionally $\beta$ increases to $>1$ so that the IBS encloses
the pulsar and a peak appears at $\phi=0.75$ in the time-averaged LAT orbital light curve.
If there were other strong evidence for episodic variability, then correlating the LAT
light curve with these states could separate the two cases.

	Evidence for modulation in the gamma-ray light curve has been claimed
in two other pulsar binaries, both with sub-stellar companions 
\citep[PSR~J1959+2048, PSR~J1311$-$3430;][]{wtch+12,xw15,arjk+17}.
For these, the modulation was strongest at high energies ($\gapp$\,GeV),
although with limited significance, this trend is not very strong.
In contrast for J2241 our best evidence for modulation is seen in the
low-energy band. Here we might expect the IBS power-law emission to
best stand out from the very hard $\Gamma=1.33$ pulsed emission. 
Unlike PSR~J1311$-$3430 we find the strongest modulation signal when
we include all spin phases in the analysis. Thus without pulse phase cuts
the low-energy spectral cut may help isolate a modulated IBS signal.
If IBS emission is dominated by synchrotron, we expect that in the plasma
rest frame the spectral peak is below $\sim 150$\, MeV, due to radiation
reaction limits. We would then require bulk Doppler boosting to shift
the IBS emission into the LAT band \citep[][]{bkk08}.
We can speculate that the bulk Doppler factor is lower for J2241 than the other two sources, leaving
the beamed IBS emission with a sub-GeV peak.
The $\Gamma\approx 2.3$ synchrotron emission implies an effective 
$p_1\approx 3.6$ index for the electrons contributing to the LAT flux. Then
with Doppler boosted modulation $\sim 2 \propto \Gamma_D^{(5+p_1)/2}$ (Figure~\ref{fig:fig2}) 
we would infer an IBS bulk $\Gamma_D \approx 1.2$ \citep[e.g.,][]{ar17}.
Inverse-Compton scattering (ICS) processes may also
be important for the higher-energy emission in other sources \citep[e.g.,][]{b14}.

	It will be especially helpful to measure the J2241 system inclination.
IBS emission is most visible at higher inclinations, as for
PSR~J1959+2048 \citep[$65^\circ$;][]{rcft+07}
and PSR~J1311$-$3430 \citep[$57-81^\circ$;][]{rfb15}.
The broad hump of the J2241 $\gamma$-ray peak suggests an intermediate inclination.
\citet[][]{kjrf+11} do not see evidence of radio eclipses or dispersion
delay in this binary, suggesting that the inclination is not
very large. Optical imaging and spectroscopy may help to constrain the
inclination $i$ value \citep[e.g.,][]{rs16,whvb+17}.
{\it Fermi} LAT will keep collecting data, and additional X-ray 
observations can test for IBS emission. While J2241 is too faint to be detected,
{\it NuSTAR} observations of brighter, hard power-law sources
might help connect these two components.

	In sum this is the third companion-evaporating
spider binary to show evidence for orbital modulation in the $\gamma$-ray
band. Interestingly all three have sub-stellar companions and
so are of the BW or Tidarren type. The RB systems with stellar mass companions 
often show strong modulation in the X-ray band. For J2241 such modulation,
if present, is much weaker. It will be interesting to examine the
rest of the spider population to see if this dichotomy persists
and to probe its physical origin. 

\bigskip
\bigskip


The \textit{Fermi}-LAT Collaboration acknowledges support for LAT development,
operation and data analysis from NASA and DOE (United States),
CEA/Irfu and IN2P3/CNRS (France), ASI and INFN (Italy), MEXT, KEK, and JAXA (Japan),
and the K.A.~Wallenberg Foundation, the Swedish Research Council
and the National Space Board (Sweden). Science analysis support
in the operations phase from INAF (Italy) and CNES (France) is also
gratefully acknowledged. This work performed in part under DOE Contract DE-AC02-76SF00515.
This research was supported by Basic Science Research Program through
the National Research Foundation of Korea (NRF)
funded by the Ministry of Science, ICT \& Future Planning (NRF-2017R1C1B2004566).
Support was also obtained from NASA under grant 80NSSC17K0024.
Work at NRL is supported by NASA.


\begin{thebibliography}{0}%
\makeatletter
\providecommand \@ifxundefined [1]{%
 \@ifx{#1\undefined}
}%
\providecommand \@ifnum [1]{%
 \ifnum #1\expandafter \@firstoftwo
 \else \expandafter \@secondoftwo
 \fi
}%
\providecommand \@ifx [1]{%
 \ifx #1\expandafter \@firstoftwo
 \else \expandafter \@secondoftwo
 \fi
}%
\providecommand \natexlab [1]{#1}%
\providecommand \enquote  [1]{``#1''}%
\providecommand \bibnamefont  [1]{#1}%
\providecommand \bibfnamefont [1]{#1}%
\providecommand \citenamefont [1]{#1}%
\providecommand \href@noop [0]{\@secondoftwo}%
\providecommand \href [0]{\begingroup \@sanitize@url \@href}%
\providecommand \@href[1]{\@@startlink{#1}\@@href}%
\providecommand \@@href[1]{\endgroup#1\@@endlink}%
\providecommand \@sanitize@url [0]{\catcode `\\12\catcode `\$12\catcode
  `\&12\catcode `\#12\catcode `\^12\catcode `\_12\catcode `\%12\relax}%
\providecommand \@@startlink[1]{}%
\providecommand \@@endlink[0]{}%
\providecommand \url  [0]{\begingroup\@sanitize@url \@url }%
\providecommand \@url [1]{\endgroup\@href {#1}{\urlprefix }}%
\providecommand \urlprefix  [0]{URL }%
\providecommand \Eprint [0]{\href }%
\providecommand \doibase [0]{http://dx.doi.org/}%
\providecommand \selectlanguage [0]{\@gobble}%
\providecommand \bibinfo  [0]{\@secondoftwo}%
\providecommand \bibfield  [0]{\@secondoftwo}%
\providecommand \translation [1]{[#1]}%
\providecommand \BibitemOpen [0]{}%
\providecommand \bibitemStop [0]{}%
\providecommand \bibitemNoStop [0]{.\EOS\space}%
\providecommand \EOS [0]{\spacefactor3000\relax}%
\providecommand \BibitemShut  [1]{\csname bibitem#1\endcsname}%
\let\auto@bib@innerbib\@empty
\end{thebibliography}%


\begin{thebibliography}{}
\expandafter\ifx\csname natexlab\endcsname\relax\def\natexlab#1{#1}\fi

\bibitem[{{Abdo} {et~al.}(2013){Abdo}, {Ajello}, {Allafort}, {Baldini},
  {Ballet}, {Barbiellini}, {Baring}, {Bastieri}, {Belfiore}, {Bellazzini}, \&
  et~al.}]{fermi2PC}
{Abdo}, A.~A., {Ajello}, M., {Allafort}, A., {et~al.} 2013, \apjs, 208, 17

\bibitem[{{Acero} {et~al.}(2015){Acero}, {Ackermann}, {Ajello}, {Albert},
  {Atwood}, {Axelsson}, {Baldini}, {Ballet}, {Barbiellini}, {Bastieri},
  {Belfiore}, {Bellazzini}, {Bissaldi}, {Blandford}, {Bloom}, {Bogart},
  {Bonino}, {Bottacini}, {Bregeon}, {Britto}, {Bruel}, {Buehler}, {Burnett},
  {Buson}, {Caliandro}, {Cameron}, {Caputo}, {Caragiulo}, {Caraveo},
  {Casandjian}, {Cavazzuti}, {Charles}, {Chaves}, {Chekhtman}, {Cheung},
  {Chiang}, {Chiaro}, {Ciprini}, {Claus}, {Cohen-Tanugi}, {Cominsky}, {Conrad},
  {Cutini}, {D{\'}Ammando}, {de Angelis}, {DeKlotz}, {de Palma}, {Desiante},
  {Digel}, {Di Venere}, {Drell}, {Dubois}, {Dumora}, {Favuzzi}, {Fegan},
  {Ferrara}, {Finke}, {Franckowiak}, {Fukazawa}, {Funk}, {Fusco}, {Gargano},
  {Gasparrini}, {Giebels}, {Giglietto}, {Giommi}, {Giordano}, {Giroletti},
  {Glanzman}, {Godfrey}, {Grenier}, {Grondin}, {Grove}, {Guillemot}, {Guiriec},
  {Hadasch}, {Harding}, {Hays}, {Hewitt}, {Hill}, {Horan}, {Iafrate}, {Jogler},
  {J{\'o}hannesson}, {Johnson}, {Johnson}, {Johnson}, {Johnson}, {Kamae},
  {Kataoka}, {Katsuta}, {Kuss}, {La Mura}, {Landriu}, {Larsson}, {Latronico},
  {Lemoine-Goumard}, {Li}, {Li}, {Longo}, {Loparco}, {Lott}, {Lovellette},
  {Lubrano}, {Madejski}, {Massaro}, {Mayer}, {Mazziotta}, {McEnery},
  {Michelson}, {Mirabal}, {Mizuno}, {Moiseev}, {Mongelli}, {Monzani},
  {Morselli}, {Moskalenko}, {Murgia}, {Nuss}, {Ohno}, {Ohsugi}, {Omodei},
  {Orienti}, {Orlando}, {Ormes}, {Paneque}, {Panetta}, {Perkins},
  {Pesce-Rollins}, {Piron}, {Pivato}, {Porter}, {Racusin}, {Rando}, {Razzano},
  {Razzaque}, {Reimer}, {Reimer}, {Reposeur}, {Rochester}, {Romani},
  {Salvetti}, {S{\'a}nchez-Conde}, {Saz Parkinson}, {Schulz}, {Siskind},
  {Smith}, {Spada}, {Spandre}, {Spinelli}, {Stephens}, {Strong}, {Suson},
  {Takahashi}, {Takahashi}, {Tanaka}, {Thayer}, {Thayer}, {Thompson},
  {Tibaldo}, {Tibolla}, {Torres}, {Torresi}, {Tosti}, {Troja}, {Van Klaveren},
  {Vianello}, {Winer}, {Wood}, {Wood}, \& {Zimmer}}]{fermi3fgl}
{Acero}, F., {Ackermann}, M., {Ajello}, M., {et~al.} 2015, \apjs, 218, 23

\bibitem[{{Acero} {et~al.}(2016){Acero}, {Ackermann}, {Ajello}, {Albert},
  {Baldini}, {Ballet}, {Barbiellini}, {Bastieri}, {Bellazzini}, {Bissaldi},
  {Bloom}, {Bonino}, {Bottacini}, {Brandt}, {Bregeon}, {Bruel}, {Buehler},
  {Buson}, {Caliandro}, {Cameron}, {Caragiulo}, {Caraveo}, {Casandjian},
  {Cavazzuti}, {Cecchi}, {Charles}, {Chekhtman}, {Chiang}, {Chiaro}, {Ciprini},
  {Claus}, {Cohen-Tanugi}, {Conrad}, {Cuoco}, {Cutini}, {D{\'}Ammando}, {de
  Angelis}, {de Palma}, {Desiante}, {Digel}, {Di Venere}, {Drell}, {Favuzzi},
  {Fegan}, {Ferrara}, {Focke}, {Franckowiak}, {Funk}, {Fusco}, {Gargano},
  {Gasparrini}, {Giglietto}, {Giordano}, {Giroletti}, {Glanzman}, {Godfrey},
  {Grenier}, {Guiriec}, {Hadasch}, {Harding}, {Hayashi}, {Hays}, {Hewitt},
  {Hill}, {Horan}, {Hou}, {Jogler}, {J{\'o}hannesson}, {Kamae}, {Kuss},
  {Landriu}, {Larsson}, {Latronico}, {Li}, {Li}, {Longo}, {Loparco},
  {Lovellette}, {Lubrano}, {Maldera}, {Malyshev}, {Manfreda}, {Martin},
  {Mayer}, {Mazziotta}, {McEnery}, {Michelson}, {Mirabal}, {Mizuno}, {Monzani},
  {Morselli}, {Nuss}, {Ohsugi}, {Omodei}, {Orienti}, {Orlando}, {Ormes},
  {Paneque}, {Pesce-Rollins}, {Piron}, {Pivato}, {Rain{\`o}}, {Rando},
  {Razzano}, {Razzaque}, {Reimer}, {Reimer}, {Remy}, {Renault},
  {S{\'a}nchez-Conde}, {Schaal}, {Schulz}, {Sgr{\`o}}, {Siskind}, {Spada},
  {Spandre}, {Spinelli}, {Strong}, {Suson}, {Tajima}, {Takahashi}, {Thayer},
  {Thompson}, {Tibaldo}, {Tinivella}, {Torres}, {Tosti}, {Troja}, {Vianello},
  {Werner}, {Wood}, {Wood}, {Zaharijas}, \& {Zimmer}}]{fermigllv06}
---. 2016, \apjs, 223, 26

\bibitem[{{Ackermann} {et~al.}(2015){Ackermann}, {Ajello}, {Albert}, {Atwood},
  {Baldini}, {Ballet}, {Barbiellini}, {Bastieri}, {Bechtol}, {Bellazzini},
  {Bissaldi}, {Blandford}, {Bloom}, {Bottacini}, {Brandt}, {Bregeon}, {Bruel},
  {Buehler}, {Buson}, {Caliandro}, {Cameron}, {Caragiulo}, {Caraveo},
  {Cavazzuti}, {Cecchi}, {Charles}, {Chekhtman}, {Chiang}, {Chiaro}, {Ciprini},
  {Claus}, {Cohen-Tanugi}, {Conrad}, {Cuoco}, {Cutini}, {D'Ammando}, {de
  Angelis}, {de Palma}, {Dermer}, {Digel}, {Silva}, {Drell}, {Favuzzi},
  {Ferrara}, {Focke}, {Franckowiak}, {Fukazawa}, {Funk}, {Fusco}, {Gargano},
  {Gasparrini}, {Germani}, {Giglietto}, {Giommi}, {Giordano}, {Giroletti},
  {Godfrey}, {Gomez-Vargas}, {Grenier}, {Guiriec}, {Gustafsson}, {Hadasch},
  {Hayashi}, {Hays}, {Hewitt}, {Ippoliti}, {Jogler}, {J{\'o}hannesson},
  {Johnson}, {Johnson}, {Kamae}, {Kataoka}, {Kn{\"o}dlseder}, {Kuss},
  {Larsson}, {Latronico}, {Li}, {Li}, {Longo}, {Loparco}, {Lott}, {Lovellette},
  {Lubrano}, {Madejski}, {Manfreda}, {Massaro}, {Mayer}, {Mazziotta},
  {McEnery}, {Michelson}, {Mitthumsiri}, {Mizuno}, {Moiseev}, {Monzani},
  {Morselli}, {Moskalenko}, {Murgia}, {Nemmen}, {Nuss}, {Ohsugi}, {Omodei},
  {Orlando}, {Ormes}, {Paneque}, {Panetta}, {Perkins}, {Pesce-Rollins},
  {Piron}, {Pivato}, {Porter}, {Rain{\`o}}, {Rando}, {Razzano}, {Razzaque},
  {Reimer}, {Reimer}, {Reposeur}, {Ritz}, {Romani}, {S{\'a}nchez-Conde},
  {Schaal}, {Schulz}, {Sgr{\`o}}, {Siskind}, {Spandre}, {Spinelli}, {Strong},
  {Suson}, {Takahashi}, {Thayer}, {Thayer}, {Tibaldo}, {Tinivella}, {Torres},
  {Tosti}, {Troja}, {Uchiyama}, {Vianello}, {Werner}, {Winer}, {Wood}, {Wood},
  {Zaharijas}, \& {Zimmer}}]{fermiiso}
{Ackermann}, M., {Ajello}, M., {Albert}, A., {et~al.} 2015, \apj, 799, 86

\bibitem[{{Ajello} {et~al.}(2017){Ajello}, {Atwood}, {Baldini}, {Ballet},
  {Barbiellini}, {Bastieri}, {Bellazzini}, {Bissaldi}, {Blandford}, {Bloom},
  {Bonino}, {Bregeon}, {Britto}, {Bruel}, {Buehler}, {Buson}, {Cameron},
  {Caputo}, {Caragiulo}, {Caraveo}, {Cavazzuti}, {Cecchi}, {Charles},
  {Chekhtman}, {Cheung}, {Chiaro}, {Ciprini}, {Cohen}, {Costantin}, {Costanza},
  {Cuoco}, {Cutini}, {D'Ammando}, {de Palma}, {Desiante}, {Digel}, {Di Lalla},
  {Di Mauro}, {Di Venere}, {Dom{\'{\i}}nguez}, {Drell}, {Dumora}, {Favuzzi},
  {Fegan}, {Ferrara}, {Fortin}, {Franckowiak}, {Fukazawa}, {Funk}, {Fusco},
  {Gargano}, {Gasparrini}, {Giglietto}, {Giommi}, {Giordano}, {Giroletti},
  {Glanzman}, {Green}, {Grenier}, {Grondin}, {Grove}, {Guillemot}, {Guiriec},
  {Harding}, {Hays}, {Hewitt}, {Horan}, {J{\'o}hannesson}, {Kensei}, {Kuss},
  {La Mura}, {Larsson}, {Latronico}, {Lemoine-Goumard}, {Li}, {Longo},
  {Loparco}, {Lott}, {Lubrano}, {Magill}, {Maldera}, {Manfreda}, {Mazziotta},
  {McEnery}, {Meyer}, {Michelson}, {Mirabal}, {Mitthumsiri}, {Mizuno},
  {Moiseev}, {Monzani}, {Morselli}, {Moskalenko}, {Negro}, {Nuss}, {Ohsugi},
  {Omodei}, {Orienti}, {Orlando}, {Palatiello}, {Paliya}, {Paneque}, {Perkins},
  {Persic}, {Pesce-Rollins}, {Piron}, {Porter}, {Principe}, {Rain{\`o}},
  {Rando}, {Razzano}, {Razzaque}, {Reimer}, {Reimer}, {Reposeur}, {Saz
  Parkinson}, {Sgr{\`o}}, {Simone}, {Siskind}, {Spada}, {Spandre}, {Spinelli},
  {Stawarz}, {Suson}, {Takahashi}, {Tak}, {Thayer}, {Thayer}, {Thompson},
  {Torres}, {Torresi}, {Troja}, {Vianello}, {Wood}, \& {Wood}}]{3fhl}
{Ajello}, M., {Atwood}, W.~B., {Baldini}, L., {et~al.} 2017, \apjs, 232, 18

\bibitem[{{Akaike}(1974)}]{aic74}
{Akaike}, H. 1974, IEEE Transactions on Automatic Control, 19, 716

\bibitem[{{An} \& {Romani}(2017)}]{ar17}
{An}, H., \& {Romani}, R.~W. 2017, \apj, 838, 145

\bibitem[{{An} {et~al.}(2017){An}, {Romani}, {Johnson}, {Kerr}, \&
  {Clark}}]{arjk+17}
{An}, H., {Romani}, R.~W., {Johnson}, T., {Kerr}, M., \& {Clark}, C.~J. 2017,
  \apj, 850, 100

\bibitem[{{Arumugasamy} {et~al.}(2015){Arumugasamy}, {Pavlov}, \&
  {Garmire}}]{apg15}
{Arumugasamy}, P., {Pavlov}, G.~G., \& {Garmire}, G.~P. 2015, \apj, 814, 90

\bibitem[{{Atwood} {et~al.}(2013){Atwood}, {Albert}, {Baldini}, {Tinivella},
  {Bregeon}, {Pesce-Rollins}, {Sgr{\`o}}, {Bruel}, {Charles}, {Drlica-Wagner},
  {Franckowiak}, {Jogler}, {Rochester}, {Usher}, {Wood}, {Cohen-Tanugi}, \&
  {S.~Zimmer for the Fermi-LAT Collaboration}}]{fermiP8}
{Atwood}, W., {Albert}, A., {Baldini}, L., {et~al.} 2013, ArXiv e-prints,
  arXiv:1303.3514

\bibitem[{{Atwood} {et~al.}(2009){Atwood}, {Abdo}, {Ackermann}, {Althouse},
  {Anderson}, {Axelsson}, {Baldini}, {Ballet}, {Band}, {Barbiellini}, \&
  et~al.}]{fermimission}
{Atwood}, W.~B., {Abdo}, A.~A., {Ackermann}, M., {et~al.} 2009, \apj, 697, 1071

\bibitem[{{Bednarek}(2014)}]{b14}
{Bednarek}, W. 2014, \aap, 561, A116

\bibitem[{{Bogovalov} {et~al.}(2008){Bogovalov}, {Khangulyan}, {Koldoba},
  {Ustyugova}, \& {Aharonian}}]{bkk08}
{Bogovalov}, S.~V., {Khangulyan}, D.~V., {Koldoba}, A.~V., {Ustyugova}, G.~V.,
  \& {Aharonian}, F.~A. 2008, \mnras, 387, 63

\bibitem[{{D'Ammando}(2010)}]{d10}
{D'Ammando}, F. 2010, The Astronomer's Telegram, 2783

\bibitem[{{de Jager} {et~al.}(1989){de Jager}, {Raubenheimer}, \&
  {Swanepoel}}]{drs89}
{de Jager}, O.~C., {Raubenheimer}, B.~C., \& {Swanepoel}, J.~W.~H. 1989, \aap,
  221, 180

\bibitem[{{Edwards} {et~al.}(2006){Edwards}, {Hobbs}, \& {Manchester}}]{ehm06}
{Edwards}, R.~T., {Hobbs}, G.~B., \& {Manchester}, R.~N. 2006, \mnras, 372,
  1549

\bibitem[{{Gentile} {et~al.}(2014){Gentile}, {Roberts}, {McLaughlin}, {Camilo},
  {Hessels}, {Kerr}, {Ransom}, {Ray}, \& {Stairs}}]{grmc+14}
{Gentile}, P.~A., {Roberts}, M.~S.~E., {McLaughlin}, M.~A., {et~al.} 2014,
  \apj, 783, 69

\bibitem[{{Huang} {et~al.}(2012){Huang}, {Kong}, {Takata}, {Hui}, {Lin}, \&
  {Cheng}}]{hkth+12}
{Huang}, R.~H.~H., {Kong}, A.~K.~H., {Takata}, J., {et~al.} 2012, \apj, 760, 92

\bibitem[{{Keith} {et~al.}(2011){Keith}, {Johnston}, {Ray}, {Ferrara}, {Saz
  Parkinson}, {{\c C}elik}, {Belfiore}, {Donato}, {Cheung}, {Abdo}, {Camilo},
  {Freire}, {Guillemot}, {Harding}, {Kramer}, {Michelson}, {Ransom}, {Romani},
  {Smith}, {Thompson}, {Weltevrede}, \& {Wood}}]{kjrf+11}
{Keith}, M.~J., {Johnston}, S., {Ray}, P.~S., {et~al.} 2011, \mnras, 414, 1292

\bibitem[{{Kerr}(2011)}]{k11}
{Kerr}, M. 2011, \apj, 732, 38

\bibitem[{{Loredo}(1992)}]{l92}
{Loredo}, T.~J. 1992, in Statistical Challenges in Modern Astronomy, ed. E.~D.
  {Feigelson} \& G.~J. {Babu}, 275--297

\bibitem[{{Luo} {et~al.}(2018){Luo}, {Ransom}, {Demorest}, {Ray}, {Stovall},
  {Jenet}, {Ellis}, {van Haasteren}, {Bachetti}, \& {NANOGrav PINT developer
  team}}]{lrdr+18}
{Luo}, J., {Ransom}, S.~M., {Demorest}, P., {et~al.} 2018, in American
  Astronomical Society Meeting Abstracts, Vol. 231, American Astronomical
  Society Meeting Abstracts 231, 453.09

\bibitem[{{Reynolds} {et~al.}(2007){Reynolds}, {Callanan}, {Fruchter},
  {Torres}, {Beer}, \& {Gibbons}}]{rcft+07}
{Reynolds}, M.~T., {Callanan}, P.~J., {Fruchter}, A.~S., {et~al.} 2007, \mnras,
  379, 1117

\bibitem[{{Roberts} {et~al.}(2015){Roberts}, {McLaughlin}, {Gentile}, {Ray},
  {Ransom}, \& {Hessels}}]{rmgr+15}
{Roberts}, M.~S.~E., {McLaughlin}, M.~A., {Gentile}, P.~A., {et~al.} 2015,
  ArXiv e-prints, arXiv:1502.07208

\bibitem[{{Romani} {et~al.}(2015){Romani}, {Filippenko}, \& {Cenko}}]{rfb15}
{Romani}, R.~W., {Filippenko}, A.~V., \& {Cenko}, S.~B. 2015, \apj, 804, 115

\bibitem[{{Romani} \& {Sanchez}(2016)}]{rs16}
{Romani}, R.~W., \& {Sanchez}, N. 2016, \apj, 828, 7

\bibitem[{{Wadiasingh} {et~al.}(2017){Wadiasingh}, {Harding}, {Venter},
  {B{\"o}ttcher}, \& {Baring}}]{whvb+17}
{Wadiasingh}, Z., {Harding}, A.~K., {Venter}, C., {B{\"o}ttcher}, M., \&
  {Baring}, M.~G. 2017, \apj, 839, 80

\bibitem[{{Wu} {et~al.}(2012){Wu}, {Takata}, {Cheng}, {Huang}, {Hui}, {Kong},
  {Tam}, \& {Wu}}]{wtch+12}
{Wu}, E.~M.~H., {Takata}, J., {Cheng}, K.~S., {et~al.} 2012, \apj, 761, 181

\bibitem[{{Xing} \& {Wang}(2015)}]{xw15}
{Xing}, Y., \& {Wang}, Z. 2015, \apjl, 804, L33

\end{thebibliography}
\end{document}